\documentclass{article}
\newcommand{\bfr}{\begin{flushright}}
\newcommand{\efr}{\end{flushright}}
 
\begin{document}
\title{Compactification of spacetime in $SU(\infty)$ Yang-Mills theory
}
\author{Kiyoshi Shiraishi\\
Institute for Nuclear Study, University of Tokyo,\\ Midori-cho, Tanashi,
Tokyo 188, Japan
}
\date{Classical and Quantum Gravity {\bf 6} (1989) pp. 2029--2034
}
\maketitle
\begin{abstract}
The compactification on a torus in $SU(\infty)$ Yang-Mills theory is
considered. A special form of the configuration of a gauge field on a
torus is examined. The vacuum energy and free energy in the presence of
fermions coupled with this background in the theory are derived and
possible symmetry breaking is investigated. 
\end{abstract}

\bigskip

Recently Floratos et al. offered $SU(\infty)$ Yang-Mills (YM) theories
\cite{1} which came from the study on membrane theories \cite{2}.  We
consider, in this paper,  the
compactification on torus in the $SU(\infty)$ YM theory. A special
form of the configuration of gauge field on torus is examined.
The vacuum energy and thermodynamic potential in the presence of
fermions coupled with the YM theory in this situation are 
derived and possible symmetry breaking is investigated.

In order that our discussion should be self-contained, we start with a
brief review of $SU(\infty)$ YM theory \cite{1}. We denote the
dimension of space-time as $D$. 
The gauge fields are given by the functions which depend on the
$D$-dimensional coordinates $x^M$ as well as the coordinates of
`sphere', $\theta$ and $\phi$;
\begin{equation}
A_M(x,\theta,\phi)=\sum_{l=1}^\infty\sum_{m=-l}^l A_M^{lm}(x)\,
Y_{lm}(\theta,\phi)
\end{equation}
where $Y$ are the spherical harmonics on $S^2$. Note that the sum
over $l$ starts with $l=1$.

The field strength is defined as
\begin{equation}
F_{MN}=\partial_MA_N-\partial_NA_M+\{A_M, A_N\}
\end{equation}
where the bracket of two functions $f$ and $g$ is defined as
\begin{equation}
\{f, g\}=\frac{\partial f}{\partial\cos\theta}
\frac{\partial g}{\partial\phi}-
\frac{\partial f}{\partial\phi}
\frac{\partial g}{\partial\cos\theta}\,.
\end{equation}

The sequential operation of the bracket satisfies the Jacobi
identity:
\begin{equation}
\{\{f, g\}, h\}+\{\{h, f\}, g\}+\{\{g, h\}, f\}=0
\end{equation}
where $f$, $g$ and $h$ are functions of $\theta$ and $\phi$.

The gauge transformation of a gauge field is given by
\begin{equation}
\delta A_M=\partial_M\omega+\{A_M, \omega\}\,.
\end{equation}
At the same time the transformation of the field strength follows
\begin{equation}
\delta F_{MN}=\{F_{MN}, \omega\}\,.
\end{equation}

The YM field equation is
\begin{equation}
D_MF^{MN}\equiv\partial_MF^{MN}+\{A_M, F^{MN}\}=0\,.
\label{eq7}
\end{equation}

For later use, we introduce the matter field $\psi(x,\theta,\phi)$ in
the `adjoint representation'. This field transforms as
\begin{equation}
\delta \psi=\{\psi, \omega\}
\label{eq8}
\end{equation}
and obeys the field equation
\begin{equation}
D_MD^M\psi-m^2\psi= 0
\end{equation}
where $m$ is the mass of the $\psi$ field. $\psi$ is assumed to have
coupling with the gauge field only.

In the analysis here a set of spherical harmonics is chosen as a basis
of generators. One can write the bracket relation as
\begin{equation}
\{ Y_{lm},
Y_{l'm'}\}=\sum_{l''m''}f_{lm}{}^{l''m''}{}_{l'm'}Y_{l''m''}
\end{equation}
where $f$ is the `structure constant'. The bracket corresponds to the
commutation for generators of the usual groups.

We can find the Cartan subalgebra in this basis: if we pick
up the spherical harmonics with $m=0$, then the following are
trivially led
\begin{equation}
\{Y_{l0}, Y_{l'0}\}=0\,.
\end{equation}

Next we consider spacetime compactification. We consider
$M^{D-1}\times S^1$ ($(D-1)$-dimensional Minkowski
spacetime$\times$circle) as the background space-time. The periodicity
with respect to the
coordinate on the circle gives rise to the `Kaluza-Klein' excited 
states \cite{3}. Furthermore, since $S^1$ is a non-simply connected
manifold, non-trivial Wilson loops can be defined on it \cite{4}. In
other words, there are vacuum expactation values of the YM field on a
torus ($S^1$) (modulo gauge transformation). They can bring about
symmetry breakdown of gauge groups in ordinary YM gauge theory
\cite{5,6,7,8}. Thus the similar mechanisms are extensively studied in
the context of multidimensional unification theory \cite{9}.

In our model, we first write out the field equation. Setting
the coordinates $x^M=(x^m,y)$, $m=0, 1, 2, \dots , D-2$, the
equation (\ref{eq7}) decomposed to:
\begin{eqnarray}
D_MF^{Mn}&=&D_mF^{mn}+ D_yF^{yn}\nonumber \\
&=&\partial_mF^{mn}+\{A_m, F^{mn}\}+\partial_yF^{yn}+\{A_y,
F^{yn}\}=0\,.
\end{eqnarray}

To obtain the equation of motion for $A_n$, we impose a 
gauge condition $\partial_MA^M=0$. Further, if we neglect the
self-coupling of YM fields $A_n$, or consider the coupling only to the
`background gauge field' $\langle A_y\rangle$ so as to get a free
field equation of motion, we obtain
\begin{equation}
\partial_m^2A^n+\partial_y^2A^n+2\{\langle A_y\rangle,
\partial_yA^n\}+\{\langle A_y\rangle,\{\langle A_y\rangle,
A^n\}\}=0\,.
\end{equation}

We consider $\langle A_y\rangle=$ constant as a usual case for
arguments for Wilson loops \cite{4}, and then the background field
strength
$\langle F_{ym}\rangle=0$ satisfies the equation of motion
$D_MF^{MN}=0$ automatically. 

Now, we consider how many degrees of freedom $\langle A_y\rangle$
possesses. For an ordinary gauge group such as $SU(N)$ the degree of
freedom is as many as the rank of the group, i.e. the dimension of
Cartan subalgebra. This is true for an arbitrary dimensional
torus. In other words: suppose $T^{a'}$ belongs to the Cartan
subalgebra. Then we can expand $\langle A_y\rangle$ as
\begin{equation}
\langle A_y\rangle=\sum_{a'}\langle A_y^{a'}\rangle T^{a'}\,.
\end{equation}
This form guarantees vanishing field strength automatically especially
on a higher-dimensional torus.

We assume $\langle A_y\rangle$ can be expanded in terms of the basis of
the Cartan subalgebra even in the $SU(\infty)$ YM theory. That is to
say,
by using components of the field, it follows
\begin{equation}
\langle A_y\rangle=\sum_{l=1}^\infty\langle A_y^{l0}\rangle
Y_{l0}(\theta,\phi)\,.
\end{equation}

In a generic case, the field equation for a component field
$A_n^{lm}$ becomes a set of simultaneous infinite number of
equations:
\begin{eqnarray}
& &(\partial_m^2+\partial_y^2)A_n^{lm}+2f_{l_10}{}^{lm}{}_{l_2m}\langle
A_y^{l_10}\rangle\partial_yA_n^{l_2m}\nonumber \\
& &\qquad\qquad+f_{l_10}{}^{lm}{}_{l'm}
f_{l_20}{}^{l'm}{}_{l_3m}\langle
A_y^{l_10}\rangle\langle
A_y^{l_20}\rangle A_n^{l_3m}=0
\end{eqnarray}
where the summations over $l_1$, $l_2$ and $l_3$ are implicit, while
the sum over $m$ is unnecessary because of the `selection rule' for
the quantum number.

To simplify the equations, we can take a new basis for $\langle
A_y\rangle$ as
\begin{eqnarray}
\langle A_y\rangle&=&\sqrt{\frac{1}{3}}\langle
A_y^{(1)}\rangle Y_{10}+\sqrt{\frac{1}{15}}\langle
A_y^{(2)}\rangle Y_{20}\nonumber \\
& &+\sum_{l=3}^\infty\langle
A_y^{(l)}\rangle\frac{1}{\sqrt{2l-1}}\left(\frac{1}{\sqrt{2l+1}}Y_{l0}-
\frac{1}{\sqrt{2l-3}}Y_{l-2,0}\right)\,.
\end{eqnarray}
In this basis, the bracket operation between $\langle A_y\rangle$ and
$A_n$ can be rewritten by
\begin{equation}
\{\langle A_y\rangle, A_n\}=\frac{\partial\langle
A_y\rangle}{\partial\cos\theta}\frac{\partial A_n}{\partial\phi}
=\sum_{l'}\sum_{lm} i m \langle A_y^{(l'+1)}\rangle
A_n^{lm}Y_{l'0}Y_{lm}\,.
\end{equation}
Thus, the use of the well-known formula for multiplication of $Y$
\cite{10}
\begin{eqnarray}
&
&Y_{l_1m_1}Y_{l_1m_2}\nonumber
\\
& &=\sum_{l_3m_3}\left\{\frac{(2l_1+1)(2l_2+1)}{4\pi
(2l_3+1)}\right\}^{1/2}(l_1m_1l_2m_2|l_3m_3)(l_10l_20|l_30)Y_{l_3m_3}
\end{eqnarray}
makes the component equations simpler. In the above expression
$(l_1m_1l_2m_2|l_3m_3)$ denotes the Clebsch-Gordon coefficient in the
standard notation.

However, for a general set of $\langle A_y^{(l)}\rangle$
we also need a diagonalization of an infinite-dimensional
(mass) matrix. In this paper, rather than giving general
discussions, we investigate the case for a specific form of
$\langle A_y\rangle$ in detail. We consider the following case:
\begin{equation}
\langle A_y^{(1)}\rangle=\frac{\theta}{L}\sqrt{4\pi}\qquad
\mbox{and}\qquad\langle A_y^{(2)}\rangle=\langle A_y^{(3)}\rangle
=\cdots=\langle A_y^{(l)}\rangle=\cdots=0
\end{equation}
where $\theta$ is a constant and $L$ is the length of the circumference
of the extraspace $S^1$. This is the only case that the mass matrix is
(already) diagonal.

Since $A_n$ can be expanded in a Fourier series 
with respect to the $S^1$ coordinate, i.e.
\begin{equation}
A_n^{lm}=\sum_{k=-\infty}^\infty A_n^{lmk}e^{i2\pi k
y/L}\qquad(0\le y < L)\,.
\end{equation}
We can make up the field equation for each excited mode:
\begin{equation}
\left(\partial_m^2-\frac{(2\pi)^2k^2}{L^2}\right)A_n^{lmk}
-2\frac{2\pi}{L} km\theta A_n^{lmk}-\frac{1}{L^2}m^2\theta^2
A_n^{lmk}=0\,.
\end{equation}
Therefore the mass square of $A_n^{lmk}$ in $(D-1)$ dimensions is given
by
\begin{equation}
\frac{1}{L^2}(2\pi k+m\theta)^2
\label{eq23}
\end{equation}
where $k$ and $m$ are integers.

Based on this mass spectrum, we can evaluate the 1-loop
vacuum energy. The vacuum energy in the $SU(\infty)$ YM theory is
seemingly anticipated to diverge because of an infinite number of
`component fields'. As for our particular model, we can first suppose
that the component fields which have the label $l\le N-1$, for a finite
integer $N$. In this situation, the number of 
corresponding generators are
\begin{equation}
\sum_{l=1}^{N-1}(2l+1)=N^2-1
\end{equation}
and the number of generators which belongs to the
Cartan subalgebra is $N-1$. These are precisely coincident with the
case of the $SU(N)$ group.

According to the usual prescription \cite{4,5,6,8}, the
1-loop vacuum energy is given formally as
\begin{eqnarray}
E_{vac}&=&-\frac{(D-2)V_{D-1}}{2(4\pi)^{(D-1)/2}}\int_0^\infty dt\,
t^{-(D-1)/2-1}\nonumber \\
& &\times\sum_{l=1}^{N-1}\sum_{m=-l}^l\sum_{k=-\infty}^\infty\exp
\left\{-t\left(\frac{2\pi}{L}\right)^2\left(k+\frac{m\theta}{2\pi}
\right)^2\right\}
\end{eqnarray}
where $V_{D-1}$ is the $(D-1)$-dimensional volume of the system.
Using
Jacobi's imaginary transformation \cite{11} and regularising $E_{vac}$
by discarding an infinity, this reduces to
\begin{equation}
E_{vac}=-\frac{(D-2)V_{D-1}L}{\pi^{D/2}L^D}\Gamma(D/2)
\sum_{k=1}^\infty\frac{1}{k^D}\left[
\frac{\sin^2(Nk\theta/2)}{\sin^2(k\theta/2)}-1\right]\,.
\end{equation}
Here finite summations have been performed. In the limit
$N\rightarrow\infty$, $E_{vac}$ diverges only at $\theta=0$ modulo
$2\pi$. This fact can be easily seen from taking a limit
$D\rightarrow\infty$. In the limit the only term with $k=1$ in the sum
remains. If we assume a
vacuum with minimum energy, the expectation value of $\theta$
is zero (mod $2\pi$). (The periodicity of $2\pi$ in $\theta$ is
explained with respect to a proper gauge transformation \cite{4,7}.)

Consequently, in the pure YM theory under the assumption of
this particular $\langle A_y\rangle$, gauge symmetry is not broken
because $\langle\theta\rangle=0$ and there appear $(N^2-1)$ massless
gauge bosons. Here we should
note that there exist many local minima in the potential, and
the number of the local minima is $N-2$ in the range $0<\theta< 2\pi$.

Next we consider the matter field coupled to the background gauge field
$\langle A_y\rangle$. For a typical example, we examine a massless
Dirac fermion field in the `adjoint representation' (recall
(\ref{eq8})). For matter fields, we can take a `twisted boundary
condition' in the circle direction. Then we obtain the Fourier
expansion of the field in the following form:
\begin{equation}
\psi_n^{lm}=\sum_{k=-\infty}^\infty \psi_n^{lmk}e^{i2\pi k
y/L+i\delta y/L}\qquad(0\le y < L)
\end{equation}
where $\delta$ is a constant which represents the `twist'. The mass
spectrum is modified as
\begin{equation}
\frac{1}{L^2}(2\pi k+m\theta+\delta)^2
\label{eq28}
\end{equation}
where $k$ and $m$ are integers.

The 1-loop vacuum energy is expressed as
\begin{eqnarray}
E_{vac}(\mbox{fermion})&=&\frac{N_F
2^{[D/2]}V_{D-1}}{2(4\pi)^{(D-1)/2}}\int_0^\infty dt\,
t^{-(D-1)/2-1}\nonumber \\ 
\times& &\sum_{l=1}^{N-1}\sum_{m=-l}^l\sum_{k=-\infty}^\infty\exp
\left\{-t\left(\frac{2\pi}{L}\right)^2\left(k+\frac{m\theta+\delta}{2\pi}
\right)^2\right\}
\end{eqnarray}
where $N_F$ is the number of fermions, and after regularisation we 
obtain
\begin{eqnarray}
E_{vac}(\mbox{fermion})&=&\frac{N_F
2^{[D/2]}V_{D-1}L}{\pi^{D/2}L^D}\Gamma(D/2)
\nonumber \\
& &\cdot\sum_{k=1}^\infty\frac{\cos(k\delta)}{k^D}\left[
\frac{\sin^2(Nk\theta/2)}{\sin^2(k\theta/2)}-1\right]\,.
\end{eqnarray}
Note the overall sign of $E_{vac}(\mbox{fermion})$.

In the case with $\delta=0$, provided that $N_F$ is enough large to
overcome the contribution from YM fields, it is possible to get the
non-vanishing vacuum gauge field expactation value at finite $N$,
even after taking the limit $N\rightarrow\infty$.
The minima of $E_{vac}(\mbox{fermion})$ are located at $\theta =2\pi
p/N$,
$p=1,\ldots, N-1$.
The lowest energy of (degenerate) vacua is then
\begin{equation}
-\frac{N_F 2^{[D/2]}V_{D-1}L}{\pi^{D/2}L^D}\Gamma(D/2)\,,
\zeta(D)
\end{equation}
where $\zeta(z)$ is the zeta function.

The vacuum energy, or the effective potential for $\theta$, has an
infinite number of degenerate minima in the limit $N\rightarrow\infty$.

Many massive fermions appear when $\theta$ is located at any minima
according to the spectrum (\ref{eq28}). On the other hand,
symmetry-breaking pattern is rather complicated in the case of finite
$N$. When $\theta=2\pi/N$, there remains only $(N-1)$ massless vector
bosons associated with the generators of the Cartan subalgebra. Thus
a symmetry breakdown such as $SU(N)\rightarrow[U(1)]^N$ is expected.
However, for general finite $N$ and for general minima of $\theta=2\pi
p/N$, we see more massless gauge bosons. For example, suppose
$N=4$ and the vacuum with $p=2$. The state with $k=1$ and $m=2$ ($l=2$
or $3$) in
the spectrum (\ref{eq23}) becomes massless. Then the resulting
symmetry can be larger than $[U(1)]^N$. If $N$ is a prime number,
this
`accidental' symmetry does not emerge in any vacuum associated
with $\theta=(2\pi/N)\times(\mbox{integer})$. If we take
$N\rightarrow\infty$, we can say that the minima of the vacuum energy
as a function of
$\theta$ are located at
every point of $2\pi Q$, where $Q$ is a rational
number, $0<Q<2\pi$

The free energy can be calculated in a similar way to obtain
$E_{vac}$ \cite{5}. The technique is the same as the one in
\cite{12}, which takes the imaginary time direction as a circle.
One finds the following expression for the free energy
$F(\mbox{fermion})$ with the fermion fields considered above:
\begin{eqnarray}
F(\mbox{fermion})&=&\frac{N_F
2^{[D/2]}V_{D-1}L}{\pi^{D/2}L^D}\Gamma(D/2)
\nonumber \\
&\cdot&\left\{\sum_{k=1}^\infty\frac{\cos(k\delta)}{k^D}\left[
\frac{\sin^2(Nk\theta/2)}{\sin^2(k\theta/2)}-1\right]\right.
\nonumber \\
& &~-\frac{N^2-1}{\beta^D}\left(1-\frac{1}{2^{D-1}}\right)\zeta(D)
\nonumber \\ 
& &~+\left.2\sum_{k=1}^\infty\sum_{n=1}^\infty
\frac{\cos(k\delta)}{(L^2k^2+\beta^2n^2)^{D/2}}\left[
\frac{\sin^2(Nk\theta/2)}{\sin^2(k\theta/2)}-1\right]\right\}
\end{eqnarray}
where $\beta$ is the inverse of temperature.

For $\delta=0$, or $\delta$ at near $0$, and sufficiently large $D$,
no phase transition is expected to occur as long as the form of
$\langle A_y\rangle$ is constrained to our ansatz. That is because what
determines the
shape of the `potential' for $\theta$ is the term with $k=1$ in the
sum.
For the case with $\delta$ takes the value near $\pi/2$ and in low
dimensions, the $k=1$ term does not necessarily dominate in the
summation, and then the shape of the potential for $\theta$ is modified
even at zero temperature; in addition, the phase transition can
take place \cite{8}.

In conclusion, we see that gauge symmetry breaking in $SU(\infty)$
YM theory is feasible under the assumption with a special form of the
configuration of the gauge field on the extra $S^1$ and in the presence
of fermion fields. We did not persue the possibility of phase
transition in the case of matter fields with a special twisted 
boundary condition on $S^1$. We want to report the effect of general
twist and the dimensionality of spacetime in an effective potential for
a simpler group such as $SU(3)$ elsewhere. For $SU(\infty)$ YM theory,
we must consider the general form of  $\langle A_y\rangle$  by 
executing a diagonalization of the infinite-dimensional mass matrix from
the beginning. Otherwise, we might miss the existence of other minima
or vacua with lower energy, as in the problem of Higgs potentials
\cite{13}. The construction of other `representation' than the
`adjoint representation' is also an interesting task.  We
hope to investigate the above subjects in relation to the vacuum
energy and spontaneous symmetry breaking.

\section*{Acknowledgements}
The author thanks S. Hirenzaki for useful comments. He also thanks A.
Nakamula for discussion and Y. Hirata for reading this manuscript.
This work is supported in part by a Grant-in-Aid for
Encouragement of Young Scientist from the Ministry of Education,
Science and Culture (\# 63790150).
The author is grateful to the Japan Society for the Promotion of
Science for the fellowship. He also thanks Iwanami F\=ujukai for
financial aid.


\end{document}